\documentclass[journal]{IEEEtran}

\usepackage{graphicx}
\usepackage{color}
\usepackage{placeins}
\usepackage{float}
\usepackage{tabularx,colortbl}
\usepackage{amssymb}
\usepackage{amsthm}
\usepackage{cite}
\usepackage{amsmath}
\usepackage{makecell}
\usepackage{subfigure}

\usepackage{caption2}

\usepackage{algorithm}
\usepackage{algpseudocode}

\theoremstyle{plain}

\theoremstyle{plain}

\IEEEoverridecommandlockouts
\begin{document}
\title{Resource Allocation for Near-Field Communications: Fundamentals, Tools, and Outlooks}

\author{Bokai~Xu, Jiayi~Zhang,~\IEEEmembership{Senior Member,~IEEE}, Hongyang~Du, Zhe~Wang, Yuanwei~Liu,~\IEEEmembership{Senior Member,~IEEE}, Dusit~Niyato,~\IEEEmembership{Fellow,~IEEE}, Bo~Ai,~\IEEEmembership{Fellow,~IEEE}, and Khaled B. Letaief,~\IEEEmembership{Fellow,~IEEE}
%
%
\thanks{B. Xu, J. Zhang, Z. Wang and B. Ai are with Beijing Jiaotong University; H. Du, and D. Niyato are with Nanyang Technological University; Y. Liu is with Queen Mary University of
London; K. B. Letaief is with Hong Kong University of Science and Technology.}
}

\maketitle
\vspace{-1.8cm}

\begin{abstract}
Extremely large-scale multiple-input-multiple-output (XL-MIMO) is a promising technology to achieve high spectral efficiency (SE) and energy efficiency (EE) in future wireless systems. The larger array aperture of XL-MIMO makes communication scenarios closer to the near-field region. Therefore, near-field resource allocation is essential in realizing the above key performance indicators (KPIs). Moreover, the overall performance of XL-MIMO systems heavily depends on the channel characteristics of the selected users, eliminating interference between users through beamforming, power control, etc. The above resource allocation issue constitutes a complex joint multi-objective optimization problem since many variables and parameters must be optimized, including the spatial degree of freedom, rate, power allocation, and transmission technique. In this article, we review the basic properties of near-field communications and focus on the corresponding “resource allocation” problems. First, we identify available resources in near-field communication systems and highlight their distinctions from far-field communications. Then, we summarize optimization tools, such as numerical techniques and machine learning methods, for addressing near-field resource allocation, emphasizing their strengths and limitations. Finally, several important research directions of near-field communications are pointed out for further investigation.
\end{abstract}

\section{Introduction}
With the advent of next-generation wireless networks beyond fifth generation (B5G) and sixth generation (6G), performance requirements will become increasingly stringent, including unprecedented data rates, high reliability, global coverage, and ultradense connectivity. Various physical antennas such as holographic multiple-input-multiple-output (MIMO), intelligent reflecting surfaces (IRS), and very large scale antennas, as well as millimeter wave, ultra-dense networks, and other new technologies, can make the above possible. Among numerous 6G technologies, the extremely large-scale MIMO (XL-MIMO) can provide high spectral efficiency (SE), high energy efficiency (EE), and reliable massive access \cite{wang2023tutorial, 10332188,[59]}. Moreover, it has evolved far-field communications into near-field communications \cite{singh2023wavefront}. 
To further optimize system performance, improve user experience, reduce energy consumption, and enhance anti-interference capability, radio resource allocation has become a core component in wireless communication networks. Additionally, due to the diversity and complexity of application scenarios and wireless environment, near-field communications resource allocation is also facing significant challenges.

In contrast to traditional far-field assumptions, 6G will have a technical trend toward higher frequency migration and active/passive antenna deployment, which is expected to have fundamentally different electromagnetic propagation. Specifically, large aperture arrays may result in a Rayleigh distance exceeding the transmission distance \cite{[68]}. 
Increasing the aperture of the antenna results in scatterers and users being located in the near-field region of the antenna array. In general, near-field channels have the following typical characteristics:

$\bullet$ \textbf{\emph{Near-field Spherical Wave Properties:}} Spherical wavefronts are more accurate at describing phase differences in the near-field. Furthermore, due to the beam focusing capabilities of near-field transmissions, we can reliably communicate with multiple users in the same angular direction at different distances, which is impossible with conventional far-field beam steering \cite{[68]}. 

$\bullet$ \textbf{\emph{Near-field Non-stationary Properties:}} Additionally, the large aperture has the non-stationarity property of only receiving signals from a relatively small area known as the visibility region (VR) in different parts of the array \cite{10040750, zhi2023performance}. 

$\bullet$ \textbf{\emph{Near-field Electromagnetic Properties:}} Compared with traditional far-field massive MIMO communications, electromagnetic waves will gain new polarization flexibility and degree of freedom, and the mutual coupling effect between antennas will be intensified.

\begin{figure*}[t]
	\centering
	\includegraphics[width=0.85\textwidth]{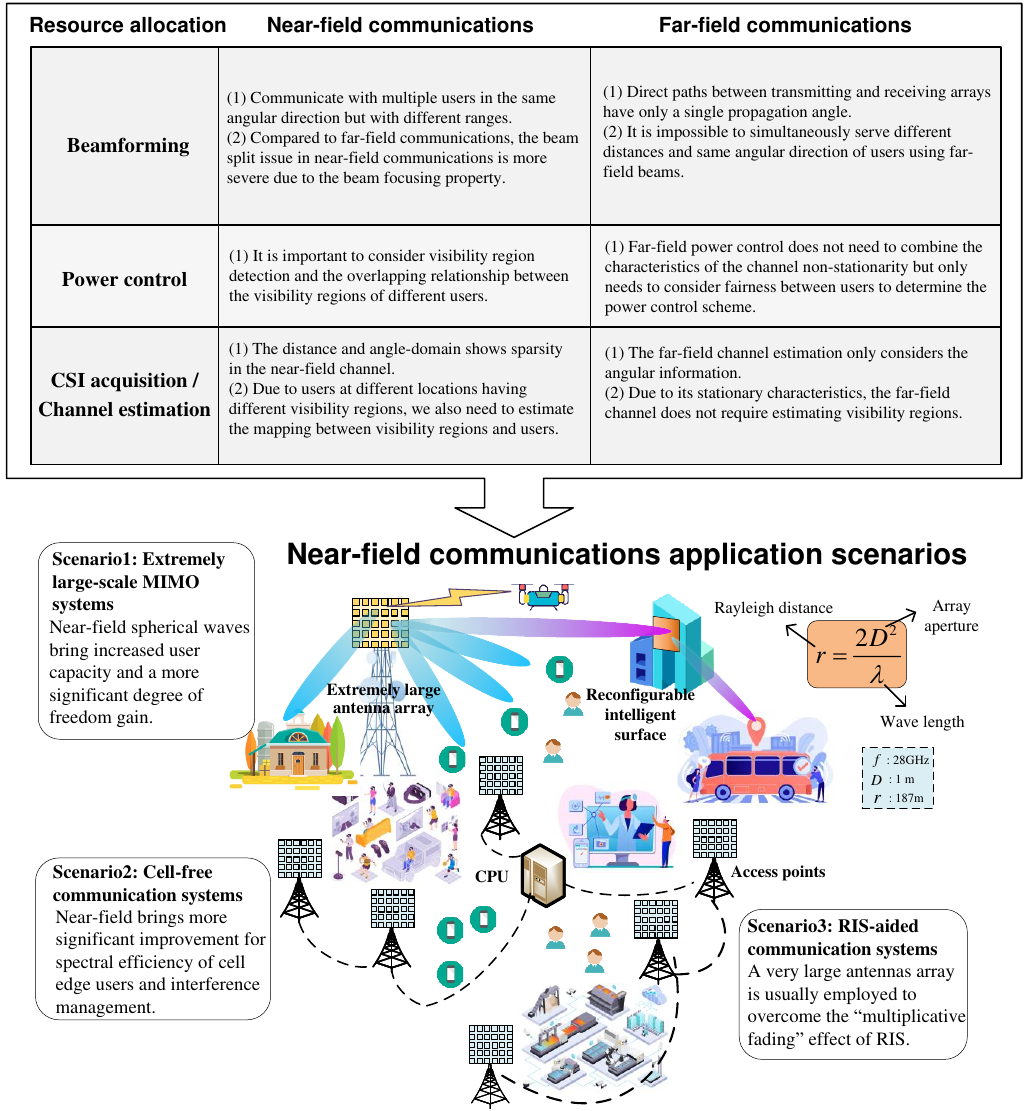}
	\caption{The major differences in resource allocation between near and far-field include beamforming, power control, and channel estimation. We consider a communication system with a frequency of $28$ GHz and an array aperture of $1$ m. Accordingly, the Rayleigh distance is $187$ m, and users are more likely located in the near-field region. The major application scenarios for near-field resource allocation include XL-MIMO systems, RIS-aided communication systems, and cell-free communication systems.} \label{Near-field communication}
	\vspace{+5pt}
\end{figure*}

Based on the above characteristics of the near-field channel, there are many resources to be allocated, including power control, beamforming, antenna selection, user scheduling, and channel estimation, mechanisms of which are different from far-field. The SE, EE, bit error rates (BER), user fairness, and quality-of-service (QoS) are common metrics for assessing far-field communications performance. Near-field communications not only improve performance in the above metrics but also bring new metrics such as mutual information, effective degrees of freedom (EDoF), beam depth (BD), etc. Near-field communications have different optimization objectives and constraints due to differences in resource allocation issues and evaluation metrics. Consequently, corresponding optimization tools and methods need to be considered. In Fig. \ref{Near-field communication}, we illustrate some typical applications of near-field communications for improving the performance of wireless systems and compare resources in near-field and far-field.

Due to the new characteristics of near-field communications, specifically for XL-MIMO, traditional far-field resource allocation schemes are no longer applicable. At the same time, many new antenna structures and signal processing architectures were introduced, so the optimization algorithms needed to be redesigned. 
To fully utilize near-field resources, joint optimization designs are often required, making the optimization solution extremely difficult and often unable to obtain a closed form solution. 
Therefore, a common strategy for addressing high-dimensional joint resource design problems involves converting non-convex optimization problems into convex optimization problems using convex relaxation, successive convex approximation (SCA), etc. On the other hand, we can also use algorithms such as alternating optimization, deep reinforcement learning (DRL), and other optimization techniques are employed to solve some non-convex optimization problems with specific constraints.

\begin{figure}
	\centering
	\subfigure[We set the number of antennas $N$ is 512, the number of paths is 20, and the carrier frequency $f_{c}$ is 30 GHz. The total number of sampling distances $S$ is 2071. Distance and angle-domain channel vectors in the near-field and far-field. The near-field channel has better sparsity in the distance and angle-domain than in the angle-domain.]{
		\begin{minipage}[b]{0.48\textwidth}
			\includegraphics[width=0.45\textwidth]{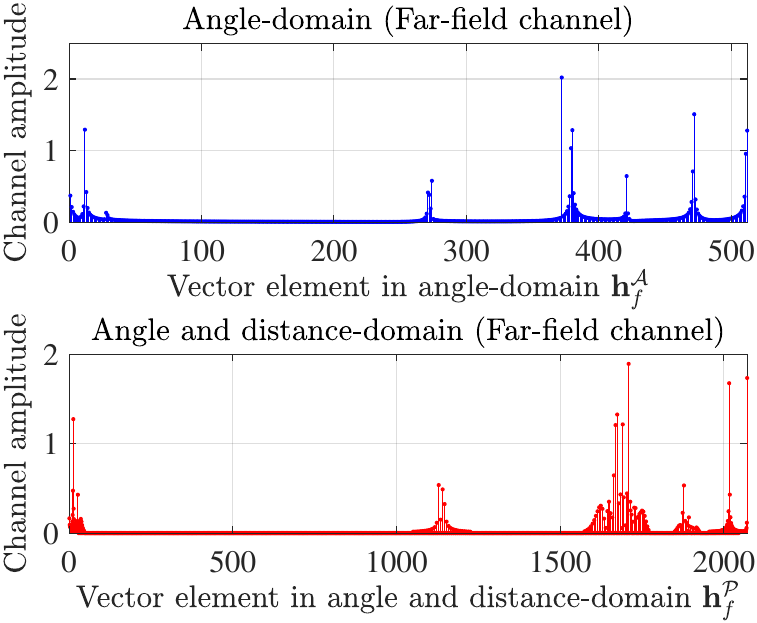} 
			\includegraphics[width=0.47\textwidth]{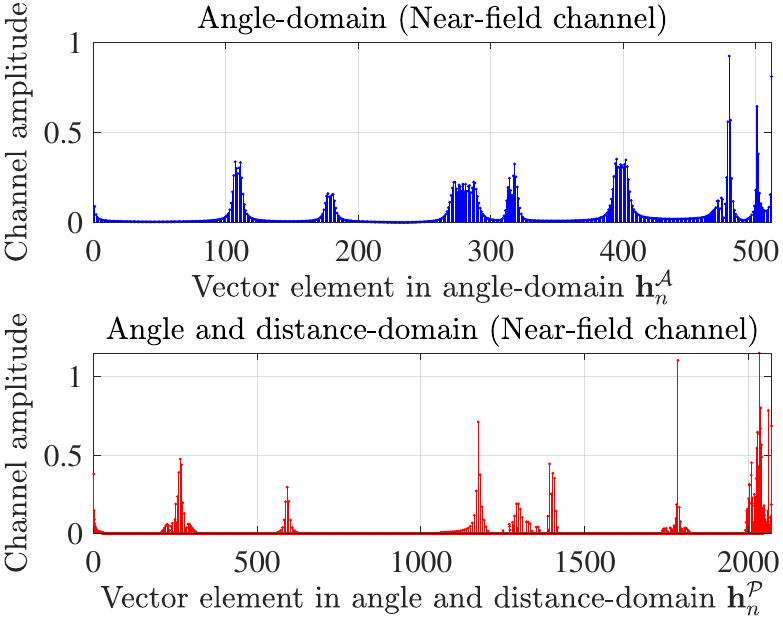}
		\end{minipage}
		\label{fig:grid_4figs_1cap_2subcap_1}
	}
    	\subfigure[The focus point changes with the frequency. The central carrier frequency is $28$ GHz and the user location is ($9$ m, $6$ m). Conventional far-field beam from a specific distance continues toward infinity, while near-field beam has a finite depth around the focal point.]{
    		\begin{minipage}[b]{0.48\textwidth}
   		 	\includegraphics[width=0.45\textwidth]{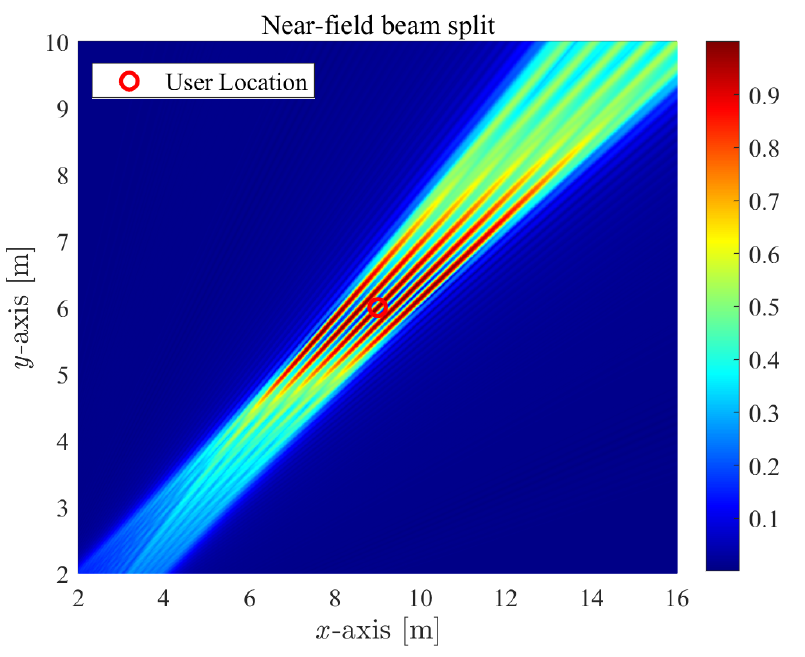}
		 	\includegraphics[width=0.45\textwidth]{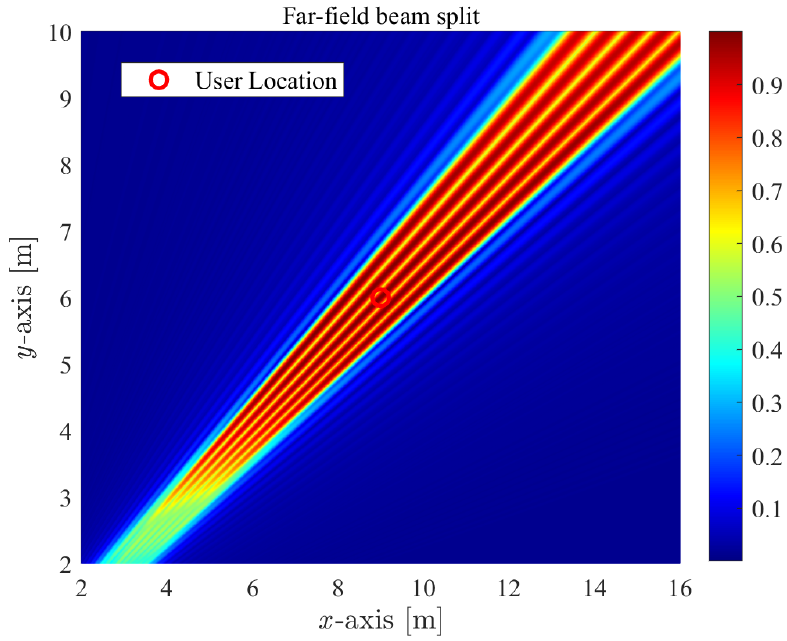}
    		\end{minipage}
		\label{fig:grid_4figs_1cap_2subcap_2}
    	}
	\caption{Comparison of near-field and far-field channel characteristics. Figure (a) compares the differences between near- and far-field from the perspective of channel sparsity. Moreover, Figure (b) compares the differences between near- and far-field from the perspective of antenna array gain at spatial positions.}
	\label{near-field channel}
\end{figure}

With the focus on near-field resource allocation, the main contributions of this paper can be summarized as follows:

$\bullet$ We discuss the channel characteristics of near-field communications. We also elucidate the disparities between near-field and far-field resource allocation. Furthermore, we explore the challenges associated with resource allocation in near-field communications and propose possible solutions.

$\bullet$ We review approaches for near-field resource allocation. The major optimization tools include traditional numerical optimization, heuristic optimization, and machine learning (ML) optimization. We then present the process of solving general near-field resource allocation problems, highlighting the strengths and limitations between traditional numerical optimization algorithms and ML algorithms. Moreover, the characteristics of the algorithm and the types of problems that each optimization tool can handle are revealed.

$\bullet$ We simulate and compare traditional optimization algorithms with ML algorithms.  Our investigated solutions can serve as a novel framework for near-field modeling and performance optimization. Finally, important research directions are highlighted for further studies.

\begin{figure*}[t]
	\centering
	\includegraphics[width=0.9\textwidth]{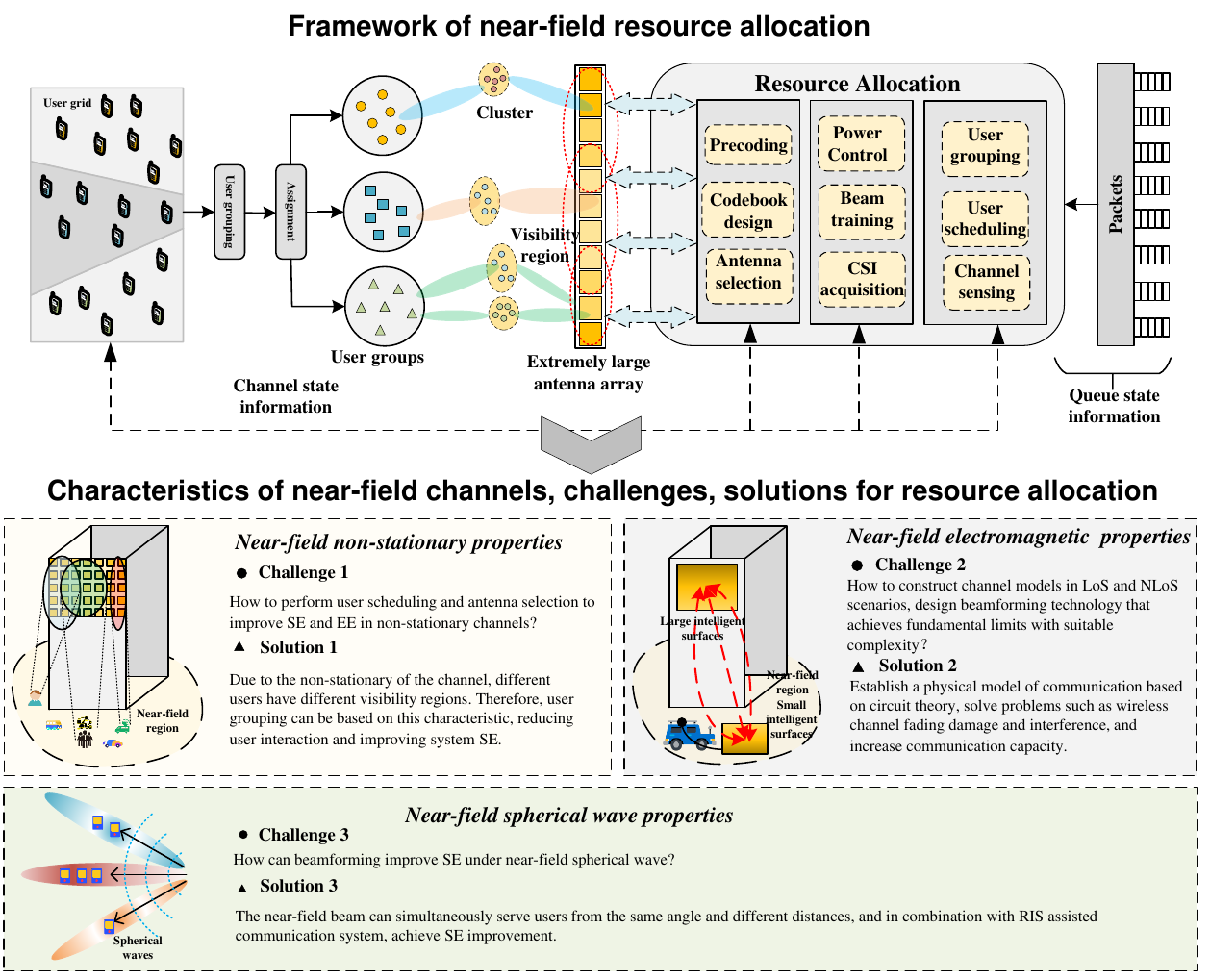}
	\caption{At the base station, the signal arrives at the receiver through resource allocation strategies such as precoding and power control, passing through the near-field channel. At the receiver, user grouping and scheduling strategies improve efficiency. Besides, we introduce three major challenges and solutions for near-field resource allocation.} \label{Fig.NearReg}
	\vspace{+5pt}
\end{figure*}

\begin{table*}[h]
\caption{The optimization tools for near-field resource allocation and solutions include numerical optimization, heuristic optimization, and machine learning.}
	\label{tab2}
	\centering
	\footnotesize
	\renewcommand{\arraystretch}{1.25}
\begin{tabular}{llllllll}
\hline\hline
\textbf{\makecell{Opimization \\algorithm}} &\textbf{\makecell{Problem \\type}} &\textbf{\makecell{Beamforming\\/Combining\\/Power control}} &\textbf{\makecell{Codebook \\design}} &\textbf{\makecell{Antenna \\Selection}} &\textbf{\makecell{Beam \\training}} & \textbf{\makecell{User grouping\\/Scheduling}} &\textbf{ Sensing}  \\ \hline \hline
\multicolumn{8}{l}{\emph{\textbf{Numerical Optimization based Approach}}}                                                                                                              \\ \hline \hline
\emph{\makecell{Alternating\\ optimization}}                     & Combinatorial optimization           &          \makecell{\space \cite{[68]}}                   &        \makecell{\space   \cite{10217152}}        &         \makecell{\space  -- }              &     \makecell{\space  \cite{10217152} }              &   \makecell{\space  -- }                &     \makecell{\space  \cite{10135096}}                       \\ \hline
\emph{\makecell{Riemannian \\manifold optimization}}                     &     \makecell{Non-convex \\constant-modulus constraint}          &     \makecell{\space  \cite{[68]}}          &           \makecell{\space  \cite{10217152}}                         &    \makecell{\space  -- }             &     \makecell{\space \cite{10217152}}           &     \makecell{\space  -- }              &     \makecell{\space  -- }               \\ \hline
       \emph{\makecell{Search algorithm}}                 &     \makecell{MINLP\\/NP-complete}         &        \makecell{\space \cite{zhi2023performance, [59]}}        &            \makecell{\space  -- }                        &          \makecell{\space  -- }             &      \makecell{\space  -- }             &      \makecell{\space  \cite{10040750}, \cite{li2023multiuser}}         &    \makecell{\space  -- }                   \\ \hline
                \emph{\makecell{Lagrange multiplier\\/SDR method}}        &    \makecell{QCQP}          & \makecell{\space \cite{[48]}}           &          \makecell{\space  -- }                          &      \makecell{\space  \cite{[48]} }                 &      \makecell{\space  -- }             &     \makecell{\space  -- }              &   \makecell{\space  \cite{10135096}}                   \\ \hline
\multicolumn{8}{l}{\emph{\textbf{Heuristic optimization based Approach}}}                                                                                                                                                   \\ \hline \hline
       \emph{ \makecell{Genetic\\algorithm}      }        &        \makecell{Combinatorial optimization}          &    \makecell{\space \cite{[48]} }         &      \makecell{\space  -- }                              &      \makecell{\space \cite{[48]} }           &      \makecell{\space  -- }           &      \makecell{\space  -- }             &    \makecell{\space  -- }           \\ \hline \hline
\multicolumn{8}{l}{\emph{\textbf{Machine Learning based Approach}}}                                                                                                                                                   \\ \hline \hline
        \emph{ \makecell{Deep learning}      }               &        \makecell{Linear inverse problems}          &    \makecell{\space  -- }             &               \makecell{\space \cite{[73]}}                   &       \makecell{\space  -- }                &    \makecell{\space \cite{[73]}}           &       \makecell{\space  -- }            &  \makecell{\space  -- }        \\ \hline
         \emph{ \makecell{Reinforcement learning\\(DDPG)}      }                &      \makecell{Discrete optimization}            &        \makecell{\space  -- }         &         \makecell{\space \cite{9610084} }                     &     \makecell{\space  -- }                  &     \makecell{\space  -- }              &     \makecell{\space  -- }              &     \makecell{\space  -- }                  \\ \hline
       \emph{ \makecell{Multi-agent \\Reinforcement learning\\(MADDPG)}      }                &      \makecell{Discrete optimization }            &    \makecell{\space \cite{10475888}}          &   \makecell{\space  -- }                                &         \makecell{\space \cite{10475888}}          &     \makecell{\space  -- }              &     \makecell{\space  -- }              &    \makecell{\space  -- }                   \\ \hline\hline     
\end{tabular}
\end{table*}
\begin{figure}[t]
	\centering
	\includegraphics[width=0.5\textwidth]{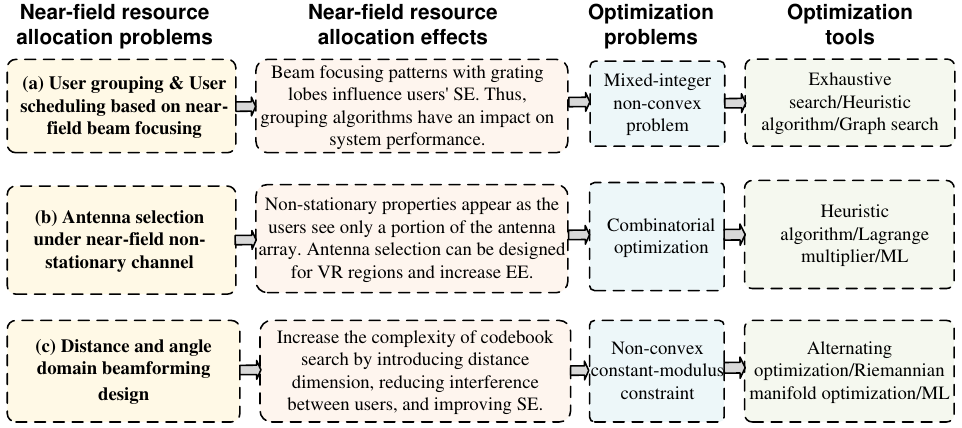}
	\caption{Near-field resource allocation problems and corresponding optimization problems and tools. (a) User grouping and scheduling based on near-field beam focusing can be modeled as a mixed-integer non-convex problem and solved using search class methods \cite{li2023multiuser}. (b) The antenna selection design based on the non-stationarity of near-field channels can be modeled as a combinatorial optimization problem and solved using methods such as ML \cite{10475888}. (c) The beamforming problem in the near-field is similar to the far-field problem. The distinction arises from channel vector variations, which can be solved using traditional alternating optimization algorithms \cite{[68]}.} \label{performance comparison}
\end{figure}


\section{Characteristics of Near-field Communications}

In this section, we present new properties of near-field communications, encompassing non-stationarity, spherical beam focusing, and electromagnetic characteristics. We then introduce the near-field resource allocation framework. Furthermore, we propose the main challenges and possible solutions to improve the performance of near-field resource allocation utilizing near-field characteristics.

\subsection{Near-field Non-stationary Properties}
XL-MIMO near-field communications exhibit a characteristic of non-stationarity. Different parts of the array can observe distinct terminals when the array size becomes very large since the energy of every terminal is focused on a specific area of the array. Due to the limited channel power of outside antennas, the VR-based channel model allows radio frequency (RF) chains to assign VR to each user dynamically. Consequently, a dynamic architecture with low complexity could provide a similar level of performance as a fully digital system, but at a lower hardware cost. This makes it possible to reduce the number of RF chains and signal processing units to reduce power consumption and improve SE by scheduling users in different VRs. A major challenge for near-field resource allocation caused by the non-stationary properties is \textbf{\emph{how to carry out user scheduling and antenna selection to improve SE and EE in non-stationary channels?}}

\subsection{Near-field Spherical Wave Properties}
As the electromagnetic wave propagation model changes from far-field plane wave to near-field spherical wave, multiple propagation angles are available between the transceiver array, which can support parallel transmission of multi-stream data and significantly improve the gain of degrees of freedom. As shown in Fig. \ref{near-field channel}, the spherical wave propagation in near-field enables beam focusing in the distance and angle-domain. By exploiting this property, the beam can be focused at a specific angle and at a specific distance. Moreover, the far-field channel is sparse in the angle-domain, but suffers energy leakage in the distance and angle-domain. The near-field channel has better sparsity in the distance and angle-domain than in the angle-domain. Regarding the near-field spherical wave property, the challenge lies in \textbf{\emph{how can beamforming improve SE, EE, and EDoF under the near-field spherical wave?}}
Capitalizing on the capability of beam focusing by optimizing the sum rate in multi-user systems through alternating optimization, the BS can naturally generate beams with spherical wavefronts to distinguish users located at similar angles but at different distances.
The near-field LoS XL-MIMO modeling yields a rank much more significant than one in the far-field, thus enabling the possibility of spatial multiplexing even in a free-space environment. This allows for improved spatial multiplexing and spatial diversity, effectively increasing the number of independent data streams that can be transmitted simultaneously. As a result, EDoF can be enhanced, enabling higher data rates and improved system capacity.

\subsection{Near-field Electromagnetic Properties}
Another new feature of near-field communications is the introduction of electromagnetic properties due to the change in the aperture of the antenna array. When the aperture of a linear array tends to be spatially continuous, i.e., holographic MIMO, the propagation mode, polarization, and spatial freedom of the electromagnetic wave in space will bring about great differences. In contrast to far-field communications, the influence of antennas on the wireless channel propagation environment is further amplified. Consequently, it becomes imperative to incorporate microwave network theory, impedance matching networks, and other methodologies to jointly model antennas and propagation channels. In addition, the compact antenna array has more sampling points in space and can capture more wavenumber domain information. By utilizing this mechanism, spatial electromagnetic waves can be exploited to achieve extreme spatial resolution and high SE. Therefore, another major challenge of near-field communications arising from these unique electromagnetic properties is \textbf{\emph{how to construct channel models for the free-space propagation environment (LoS) and the arbitrary scattered propagation environment (NLoS) scenarios?}} The LoS channel modeling can be studied by two major models based on different modeling concepts: Green's function-based modeling and complex channel response representation. The Green's function based method involves the numerical solution of Maxwell's equations, allowing it to accurately capture the electromagnetic (EM) characteristics of the channel. This method is suitable for both discrete and continuous apertures.
On the other hand, when modeling the NLoS propagation channel, two major methods can be employed, taking into account the scattering mechanisms involved. These methods are the Fourier plane-wave representation and the array response vector representation. The Fourier plane-wave representation method is applicable to arbitrary scattering mechanisms and can be utilized for both discrete and continuous apertures. It encompasses the source, receiver, and angular responses within its framework. In addition, we need to design corresponding beamforming techniques to reach the fundamental limits with an acceptable complexity based on the above channel models.

In Fig. \ref{Fig.NearReg}, we summarize the framework for near-field resource allocation and propose three challenges and possible solutions.

\section{Optimization Tools for Resource Allocation in Near-Field Communications}
In this section, we delve into optimization tools for near-field resource allocation and classify them into three distinct categories: numerical optimization based methods, heuristic optimization based methods, and ML based methods. Subsequently, we explore the optimization problems to which these algorithms can be applied, examine resource allocation scenarios, and analyze the key characteristics of these algorithms. 
Due to the ability of near-field beam focusing to provide effective resolution in the distance dimension, we can utilize this feature for more effective user grouping and scheduling. For the antenna selection problem under near-field non-stationary channel characteristics, using subarray selection can achieve more efficient and distributed processing, allowing the system to process more extensive and complex datasets without affecting performance and accuracy. In addition, we can fully utilize the additional spatial degrees of freedom introduced by the near-field spherical wave model by combining adjustable RF chains, encrypted arrays, and other methods to design corresponding beamforming algorithms to improve system gain. We have summarized the corresponding optimization problems and methods. As shown in Fig. \ref{performance comparison}, we summarize the near-field resource allocation problems and corresponding suitable optimization methods.
\subsection{Numerical Optimization based Approach}
Due to its reliability, versatility, and solid mathematical foundation, traditional numerical optimization algorithms become valuable tools for solving resource allocation problems and have found widespread applications in telecommunications. 
Due to the unique nature of resource allocation problems and the diverse structures of antenna arrays, new optimization problems have to be formulated. Depending on the specific forms of the objectives and constraints, one can select appropriate numerical optimization tools to address them.
\begin{figure*}[t]
	\centering
	\includegraphics[width=0.75\textwidth]{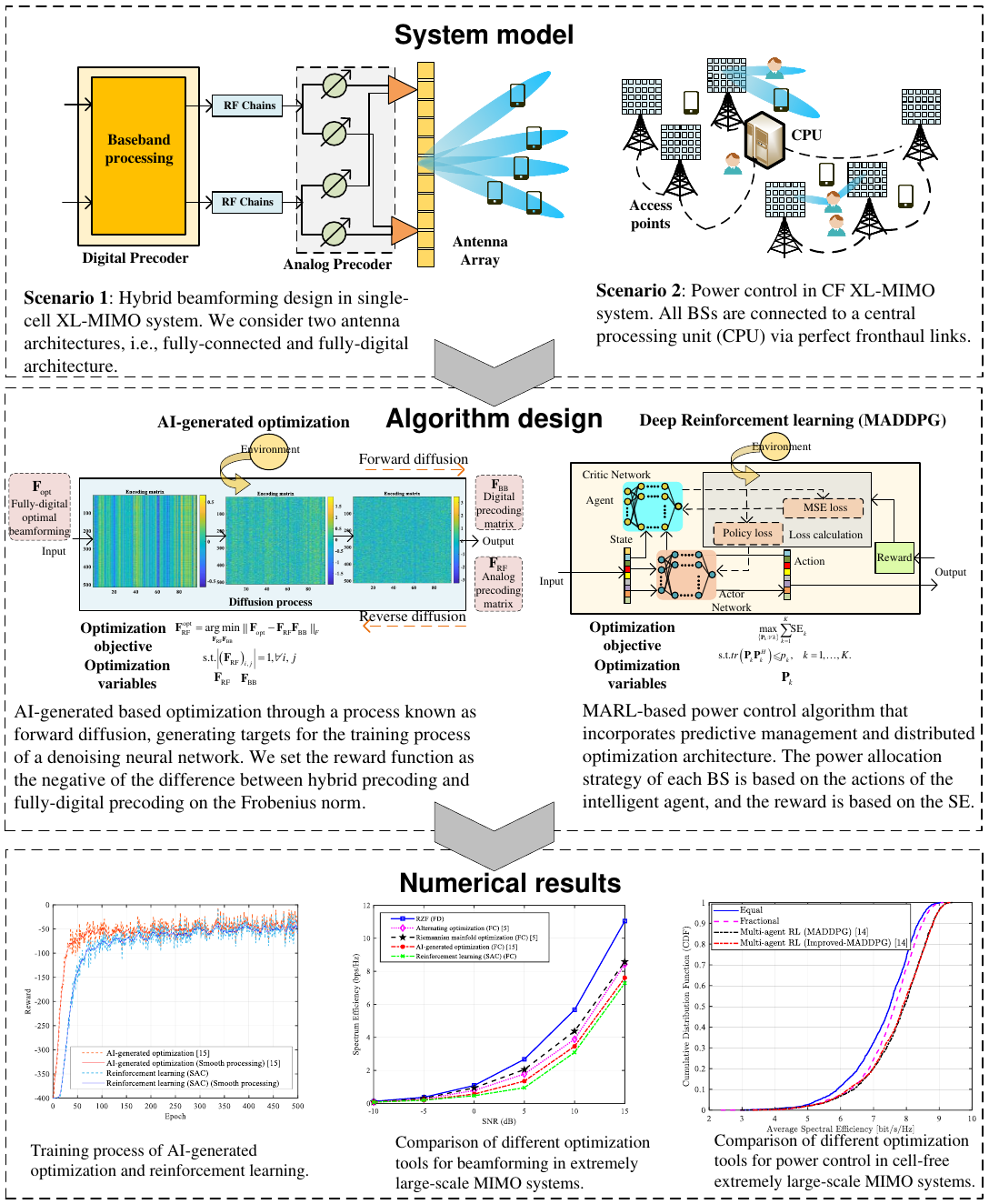}
	\caption{\textbf{Scenario 1:} Beamforming design in XL-MIMO systems. The BS has $N = 512$ antennas and the carrier frequency is $f = 28$ GHz. The number of RF chains is $4$, serving $8$ receivers. \textbf{Scenario 2:} Power control in cell-free communication systems. The number of access points is $9$, serving $6$ receivers. The communication bandwidth is $20$ MHz  and the noise power is $\sigma^{2}=-69$ dBm. All transmitters transmit with a transmission power of no more than $200$ mW.} \label{simulation}
	\vspace{+5pt}
\end{figure*}

\emph{1) Alternating Optimization:} 
One type of alternating optimization is the Alternating Minimization (AM) algorithm, commonly used in collaborative filtering and matrix factorization problems. 
In contrast to the traditional far-field beam steering that can only point in a specific direction, the near-field spherical wave propagation brings the possibility of focusing the beam at a particular position. Furthermore, near-field beamforming can be transformed into a combinatorial optimization problem and solved using the AM algorithm. According to \cite{[68]}, the authors proposed an alternating design algorithm to jointly optimize the dynamic metasurface antennas (DMAs) configuration and beamforming, i.e., configurable weights of the DMAs are optimized first, and digital precoding vectors are later. Additionally, beam focusing based on the proposed algorithms can achieve notable gains in SE compared to designs assuming conventional far-field operation.  More significantly, it is shown that the SE of DMAs is comparable to that of fully digital architectures.

\emph{2) Riemannian manifold:} Optimization over a Riemannian manifold is locally analogous to optimization over the Euclidean space with smooth constraints. Therefore, a well-developed conjugate gradient algorithm in Euclidean spaces can also be applied to Riemannian manifolds. Compared with far-field beam training, near-field beam training needs to search for angle parameters and additional distance-related factors to accurately focus the beam, which leads to excessive training overhead and computational complexity. The authors in \cite{10217152} modeled it as a combinatorial optimization problem with non-convex constraints in constant mode and applied the manifold optimization method to solve it. Despite slight beamforming performance degradation, training overhead can be reduced by over $99\%$ compared with the bottom-layer overall codewords exhaustive searching.

\emph{3) Search Algorithm:} 
Near-field resource allocation search schemes include greedy search, graph search, graph matching, etc. For instance, the authors in \cite{li2023multiuser} designed an efficient greedy algorithm to solve mixed integer non-convex optimization problems in near-field user grouping and antenna selection. Although this algorithm has high complexity, it can significantly improve system spectral efficiency when users are densely distributed. Moreover, an adjustable RF chain beamforming structure was proposed in \cite{[59]} to fully utilize the additional spatial degrees of freedom introduced by the near-field spherical wave to enhance the capacity of a single-user system. In addition, they proposed a low-complexity algorithm to optimize the selection matrix, which approximates exhaustive searching.

\emph{4) Other Approaches:} We illustrate additional optimization algorithms employed in near-field resource allocation. Near-field communications use a high degree of freedom and spherical wave introduced by the very large scale antenna to meet various QoS requirements of users and maintain their fairness. The problem is usually modeled as a non-convex joint high-dimensional resource optimization problem, which cannot be solved by traditional convex optimization methods. However, the problem can be transformed into a convex optimization problem by stochastic successive convex approximation theory, and then solved by using existing convex optimization tools. Additionally, there are combinatorial optimization problems related to near-field resource allocation, such as quadratically constrained quadratic program (QCQP) problems. Due to the large antenna array size, traditional semidefinite relaxation (SDR) optimization methods require an enormous computation burden, which is no longer applicable. Therefore, this necessitates low-complexity strategies, such as the constrained least square (LS) method or ML optimization tools.

\subsection{Heuristic Optimization based Approach}
Heuristic optimization based methods cover a variety of algorithms for solving complex non-convex and combinatorial optimization problems. 
One widely adopted metaheuristic approach for solving the resource allocation problem in antenna selection is the Genetic Algorithms (GA). This technique incorporates multiple search phases to effectively explore the feasible solution space and exploit the favorable properties of candidate solutions, aiming to identify promising regions within the feasible subspaces. Unlike exact optimization methods, GA does not require convex objective functions or constraints. Moreover, the computational complexity can be adjusted to match the available computational resources by tuning the input parameters and the number of iterations. However, it should be noted that, similar to other meta-heuristics, GA does not guarantee to find the optimal solution \cite{[48]}. GA is also not good with problems with constraints, and it may converge very slowly.

\subsection{Machine Learning based Approach}
ML has emerged as one of the most dynamic areas in contemporary signal processing. 
In contrast to the traditional iterative methods, where an algorithm is executed over multiple iterations, the concept of deep unfolding represents these iterations as a sequence of identical processing layers. 
 
$\bullet$ \emph{Deep learning based approach}:
Compared to far-field channels, the near-field channel does not exhibit sparsity in the angle-domain but exhibits sparsity in the distance and angle-domain. Therefore, codebook design must consider both the angle and distance between the transmitter and the receiver, which differs from the far-field domain. The authors in \cite{[73]} proposed a deep learning-based beam training technique using two neural networks to estimate the optimal angle and distance for near-field beams. Moreover, the improved scheme can increase effective achievable rates and reduce the pilot overhead by approximately $95\%$ compared to the beam sweeping technique.

$\bullet$ \emph{Deep reinforcement learning based approach}:
For high-dimensional resource allocation problems such as time, space, frequency, and degree of freedom in near-field, traditional numerical optimization algorithms often cannot solve this kind of joint optimization design problem. The joint design method based on reinforcement learning algorithm can dynamically adjust its own decision-making strategy through the interaction between the agent and the environment, and obtain the optimal expected reward.
The authors in \cite{9610084} proposed a framework to optimize the codebook beam patterns based on the environment. Furthermore, the proposed solution produces beams with similar SNR and approaches ideal beam shapes. For near-field antenna selection and power control problems, single-agent reinforcement learning is no longer suitable for such complex scenarios when the user has mobility or is served by multiple base stations. In contrast, multi-agent reinforcement learning performs better in this combinatorial optimization problem \cite{10475888}. Moreover, it can reduce CSI interaction, reduce complexity and improve energy efficiency by using part of information to exchange and share information among users.

To sum up, we present a list of main resource allocation problems and optimization tools in Table \ref{tab2} for ease of reference.

\section{Use Cases}
This section evaluates and demonstrates near-field resource allocation designs and optimization. We consider two scenarios for near-field resource allocation: XL-MIMO and cell-free communication systems.

\textbf{Scenario 1:} 
We consider two antenna architectures, i.e., fully-connected and fully-digital precoded architectures. We set the number of RF chains for a fully-connected antenna architecture to 4.
For performance comparison of beamforming schemes, we consider the three traditional numerical optimization algorithms: \emph{Optimal beamforming}, \emph{Riemannian manifold optimization}, \emph{alternating optimization} \cite{[68]} and \emph{reinforcement learning}. Moreover, we compare a generative AI-based method, called \emph{AI-generated optimization}, whose key technology is generative diffusion models \cite{du2023beyond}. Through a process known as forward diffusion, the generative diffusion model gradually adds Gaussian noise over time, generating targets for  the training process of a denoising neural network. From Fig. \ref{simulation}, we can observe a trade-off between performance and algorithm complexity: although traditional Riemannian optimization schemes can provide the performance of fully-digital structures more closely, their complexity is too high. In contrast, alternating optimization can achieve a balance between complexity and performance. Although AI-based methods require computational power for training, utilizing the generalization ability of neural networks can better adapt to changes in dynamic environments. Furthermore, the above near-field beamforming schemes can reliably communicate with multiple users in the same angular direction but with different ranges, which is impossible in far-field beam steering. In addition, if the receivers have high mobility, i.e., the channel is constantly changing, the two methods require recalculation at every moment, which is extremely time-consuming. Alternately, AI-generated optimization is more suitable for complex and ever-changing scenarios. However, performance is slightly reduced. Moreover, AI-generated optimization is superior to the reinforcement learning algorithm, e.g., Soft Actor Criticism (SAC), which has faster convergence and better performance.

\textbf{Scenario 2:} As shown in Fig. \ref{simulation}, we consider the near-field power control scenario of multiple base stations. It is observed that the \emph{multi-agent reinforcement learning (MADDPG)} \cite{10475888} is better than the traditional power control scheme. 
Traditional power control algorithms cannot effectively utilize historical information for real-time CSI acquisition and management. On the contrary, the MADDPG-based multi-agent reinforcement learning algorithm (MARL) utilizes decisions and interaction among multiple agents to better solve complex optimization problems in the near-field.
To ensure adaptability to the changing wireless conditions, we have integrated an online learning mechanism into the network architecture. This mechanism enables agents to continuously learn and update their policies based on the prevailing wireless conditions. Consequently, the trained MARL model remains up-to-date in the dynamic wireless environment, leading to superior system performance.
Moreover, the training time of MADDPG and Improved-MADDPG algorithms is $0.554$ s and $0.610$ s respectively, and the convergence time is $178$ s and $75$ s respectively. Furthermore, the Improved-MADDPG algorithm can significantly improve the convergence speed of the MADDPG problem, and compared to traditional optimization algorithms, multi-agent deep reinforcement learning algorithms can significantly reduce interference between antennas and achieve better power control performance.

\section{Conclusions and Future Directions}
In this article, we provided a comprehensive overview
of near-field resource allocation for XL-MIMO systems. Specifically, we first presented near-field channel characteristics. For near-field optimization problems, we presented a solution framework and optimization tools. Additionally, we explained which kinds of resource allocation problems that are suitable for numerical, heuristics, and ML.
Some other directions for future research in near-field communications are outlined as follows.

\textbf{Electromagnetic information theory:} 
Electromagnetic information theory is a promising direction, based on extremely large antenna array (ELAA) technology and holographic massive MIMO technology, to consider how to surpass the classic Massive MIMO problem. Nevertheless, this introduces high-dimensional channels, space, degrees of freedom, and other resources in the near-field and makes channel modelling, network deployment, and scheduling users more challenging. In future electromagnetic information theory research, how to allocate the resources above in the near-field will be crucial.



\textbf{Semantic Communications:} 
Semantic communication is considered a breakthrough beyond the Shannon paradigm to transmit the semantic information the source conveys successfully. 
By utilizing semantic communication, near-field communication can reduce the amount of data that needs to be transmitted, thereby minimizing bandwidth requirements and saving energy. This method utilizes the inherent semantic structure of data, surpassing traditional compression techniques, thereby achieving more effective communication.
Furthermore, near-field resource allocation enables a greater spatial degree of freedom and SE, which can accommodate new wireless communication applications requiring larger capacity.

\bibliographystyle{IEEEtran}

\bibliography{IEEEabrv,ref}
\end{document}